\documentclass[%
 reprint,
superscriptaddress,
 amsmath,amssymb,
aps,
prb,
]{revtex4-2}

\usepackage{amsmath}
\usepackage[utf8]{inputenc}
\usepackage{siunitx} 
\usepackage{xcolor}
\usepackage{natbib}
\usepackage{graphicx}
\usepackage{braket}
\usepackage{comment}
\usepackage{bm}
\usepackage{epstopdf}
\usepackage{multirow}
\usepackage{array}

\usepackage[english]{babel}
\usepackage{graphicx}
\usepackage{dcolumn}
\usepackage{bm}
\usepackage{color}
\usepackage{todonotes}
\usepackage{physics}
\usepackage{calrsfs}
\DeclareMathAlphabet{\pazocal}{OMS}{zplm}{m}{n}

\usepackage{graphicx}
\usepackage{dcolumn}
\usepackage{bm}
\usepackage{comment}

\usepackage{braket}
\usepackage{color}
\usepackage{array}
\newcolumntype{?}{!{\vrule width 1pt}}

\begin{document}

\title{{\color{black}Coulomb correlated multi-particle states of weakly confining GaAs quantum dots}}

\date{\today}

\author{Petr Klenovsk\'{y}}%
    \email{klenovsky@physics.muni.cz}
    \affiliation{Department of Condensed Matter Physics, Faculty of Science, Masaryk University, Kotl\'a\v{r}sk\'a~267/2, 61137~Brno, Czech~Republic}
    \affiliation{Czech Metrology Institute, Okru\v{z}n\'i 31, 63800~Brno, Czech~Republic}

\begin{abstract} 
%

We compute the electronic and emission properties of Coulomb–correlated multi-particle states (X$^0$, X$^\pm$, XX) in weakly confining GaAs/AlGaAs quantum dots using an 8-band $\mathbf{k}\!\cdot\!\mathbf{p}$ model coupled to continuum elasticity and configuration interaction (CI). We evaluate polarization-resolved oscillator strengths and radiative rates both in the dipole approximation (DA) and in a quasi-electrostatic beyond-dipole (BDA) longitudinal formulation implemented via a Poisson reformulation exactly equivalent to the dyadic Green-tensor kernel. For the dots studied, BDA yields lifetimes in quantitative agreement with experiment, e.g., $\tau^X=0.279\,\mathrm{ns}$ vs $0.267\,\mathrm{ns}$ (exp.) and $\tau^{XX}=0.101\,\mathrm{ns}$ vs $0.115\,\mathrm{ns}$ (exp.). The framework also reproduces electric-field tuning of the multi-particle electronic structure and emission---including the indistinguishability inferred from $P=\tau^X/(\tau^X+\tau^{XX})$---and we assess sensitivity to CI-basis size and to electron–electron and hole–hole exchange.
%


\end{abstract}


\maketitle

%
\section{Introduction}
\label{sec:intro}
Among the key components in quantum networks~\cite{Kimble2008}, quantum light sources are of dominant importance. As one of those, quantum dots (QDs) have been identified as
one of among the leading solid-state quantum light emitters~\cite{Aharonovich2016,Senellart2017,zhou2023epitaxial,Fox2025}.
Since their discovery~\cite{Ekimov1981,Ekimov1985,Leonard1993,Wegner2024} a considerable progress was obtained by improving the material quality to reduce charge noise~\cite{Kuhlmann2015,Lodahl2022}, by integrating QDs in photonic structures~\cite{Lodahl2015,Senellart2017,Liu2019,Wang2019b,Tomm2021}, by tailoring the QD properties through external electric~\cite{Bennett2010a}, magnetic~\cite{Bayer2002}, and elastic fields~\cite{Oyoko2001,seidl2006effect,Singh2010c,Gong2011e,Martin-Sanchez2018,Gaur2025}, and by implementing advanced excitation schemes~\cite{Wang2019b,Sbresny2022}. 

Along the experimental development, theoretical computational models were also improved~\cite{brasken2000full,baer2005optical,Bester2006,tomic2009excitonic,Schliwa:09,Mittelstadt2022}, in order to capture the detailed physics of QDs and guide experimental efforts. In principle, such models could be used to design QDs with tailored properties without the need to perform many resource-intensive growth and measurements. If such models are quantitatively validated, they might enable the development of quantum light sources with increasing complexity.

One of the possibilities to prepare quantum light photons is the biexciton-exciton cascade~\cite{Winik2017,Kettler2016,He2016,Ozfidan2015,Huber2018a,Lehner2023}. Clearly, a model that would correctly predict the energy ordering of the biexciton (XX) with respect to the exciton (X) would be beneficial. It should also find the correct energies of the negative trions (X$^-$) and positive trions (X$^+$) relative to X, as well as the emission rates of all of the aforementioned complexes. Clearly, it is crucial to test such a theory with an experimentally reliably measured quantum system for which complete experimental data on multiple features of the system are available~\cite{yuan_xueyong_2023_7748664}. To this end, GaAs QDs in AlGaAs nanoholes~\cite{Rastelli2004,Wang2009,Plumhof2010,Plumhof2013,Huo2013a,Yuan2018a,Huang2021a,Heyn2010,Lobl2019} are chosen in this work. The reason is their high ensemble homogeneity~\cite{DaSilva2021,Keil2017a,Rastelli2004a}, negligible built-in strain, and limited intermixing between the GaAs core and AlGaAs barriers~\cite{Zaporski2023}.
In addition, these dots also exhibit the effect of weak confinement~\cite{Zhu2024,Stobbe2012,Tighineanu2016}, considerably decreasing the radiative emission lifetime of the emitted exciton and other complexes~\cite{Reindl2019}.

Although realistic models have been applied to this system in the past, such as for GaAs/AlGaAs QDs~\cite{Wang2009}, theoretical predictions have unfortunately not yet been able to faithfully reproduce the experimentally observed values. This holds even when realistic QD structural properties and advanced theoretical models were employed~\cite{Bester2006}.

In this work, we present correlated multi-particle calculations for large GaAs/AlGaAs QDs that successfully replicate the electronic and emission properties of the system. Our analysis demonstrates that, to achieve accurate agreement with the experimental data, it is essential to account for the weak confinement effects present in these QDs.




\section{Theory model}
\label{sec:teorDesc}
\subsection{Single-particle states}
\label{subsec:kp}
In the calculations, we first implement the 3D QD model structure (size, shape, chemical composition). This is followed by the calculation of elastic strain by minimizing the total strain energy in the structure and subsequent evaluation of piezoelectricity up to non-linear terms~\cite{Bester:06,Beya-Wakata2011,Klenovsky2018}. The resulting strain and polarization fields then enter the eight-band $\mathbf{k}\!\cdot\!\mathbf{p}$ Hamiltonian~\cite{Bahder1990}.

In $\mathbf{k}\!\cdot\!\mathbf{p}$, implemented within the Nextnano++ computational suite~\cite{Birner2007}, we consider the single-particle states as linear combinations of $s$-orbital~like and $x$,~$y$,~$z$~$p$-orbital~like Bloch waves~\cite{Bahder1990,Birner2007} at $\Gamma$ point of the Brillouin zone,~i.e.,
\begin{equation}
    \psi_{a_n}(\mathbf{r}) = \sum_{\nu\in\{s,x,y,z\}\otimes \{\uparrow,\downarrow\}} \chi_{a_n,\nu}(\mathbf{r})u^{\Gamma}_{\nu}\,,
\end{equation}
where $u^{\Gamma}_{\nu}$ is the Bloch wavefunction of $s$- and $p$-like conduction and valence bands at $\Gamma$ point, respectively, $\uparrow$/$\downarrow$ marks the spin, and $\chi_{a_n,\nu}$ is~the~envelope function for $a_n \in \{ e_n, h_n \}$ [$e$ ($h$) refers to electron (hole)] of the $n$-th single-particle state.
Thereafter, the following envelope-function $\mathbf{k}\!\cdot\!\mathbf{p}$ Schr\"{o}dinger equation is solved
\begin{equation}
\label{eq:EAkp}
\begin{split}    
    &\sum_{\nu\in\{s,x,y,z\}\otimes \{\uparrow,\downarrow\}}\Bigg(\Bigg[E_\nu^{\Gamma}-\frac{\hbar^2{\bf \nabla}^2}{2m_0}+V_{0}({\bf r})\Bigg]\delta_{\nu'\nu}+\\
    &+\frac{\hbar}{2 m_0}\{\nabla,\mathbf p_{\nu'\nu}\}+ \hat{H}^{\rm str}_{\nu'\nu}({\bf r})+\hat{H}^{\rm so}_{\nu'\nu}({\bf r})\Bigg)\chi_{a_n,\nu}({\bf r})=\\
    &=\mathcal{E}^{k\cdot p}_n\cdot \chi_{a_n,\nu'}({\bf r}),
\end{split}    
\end{equation}
where the term in round brackets on the left side of the equation is the envelope function $\mathbf{k}\!\cdot\!\mathbf{p}$ Hamiltonian $\hat{H}_0^{k\cdot p}$, and $\mathcal{E}^{k\cdot p}_n$ on the right side is the $n$-th single-particle eigenenergy. Note that we use in Eq.~\eqref{eq:EAkp} the symmetrized gradient–momentum operator $\frac{\hbar}{2m_0}\{\nabla,\mathbf p\}$, which guarantees a Hermitian $\mathbf{k}\!\cdot\!\mathbf{p}$ Hamiltonian. Furthermore, $E_\nu^{\Gamma}$ is the energy of bulk $\Gamma$-point Bloch band $\nu$, $V_0({\bf r})$ is the scalar potential (e.g. due to piezoelectricity), $\hat{H}^{\rm str}_{\nu'\nu}({\bf r})$ is the Pikus-Bir Hamiltonian introducing the effect of elastic strain~\cite{Bahder1990,Birner2007,t_zibold}, and $\hat{H}^{\rm so}_{\nu'\nu}({\bf r})$ is the spin-orbit Hamiltonian~\cite{Bahder1990,t_zibold}. Further, $\hbar$ is the reduced Planck's constant, $m_0$ the free electron mass, $\delta$ the Kronecker delta, and $\nabla := \left( \frac{\partial}{\partial x}, \frac{\partial}{\partial y}, \frac{\partial}{\partial z} \right)^T$.

Furthermore, in the eight-band $\mathbf{k}\!\cdot\!\mathbf{p}$ model, the spin–orbit interaction is explicitly included through the coupling between conduction and valence bands. In particular, the valence band states are described within the total angular momentum basis $\ket{J, m_J}$ with $J = 3/2$ (heavy and light holes) and $J = 1/2$ (split-off band), where $m_J$ combines both spin and orbital angular momentum. As a result, the single-particle states $\psi_k^{(e)}$ and $\psi_l^{(h)}$ obtained from the $\mathbf{k}\!\cdot\!\mathbf{p}$ Hamiltonian represent mixed spin–orbital character. Consequently, spin is not a good quantum number in this basis and cannot be unambiguously separated or assigned to the single-particle orbitals used in subsequent configuration interaction (CI) calculations.

The aforementioned Schr\"{o}dinger equation is then solved self-consistently with the Poisson equation to improve the spatial position of electron and hole wavefunctions~\cite{Birner2007}. Note that the Poisson equation solver used in the single-particle calculations does not include Coulomb exchange.

\subsection{Configuration interaction}
\label{subsec:CI}

The single-particle states computed by the aforementioned $\mathbf{k}\!\cdot\!\mathbf{p}$ are used as basis states for CI~\cite{Bryant1987,Schliwa:09,Klenovsky2017}. In CI we consider the multi-particle ($M$) $m$-th state as
\begin{equation}
\label{eq:SDgeneralForm}
\begin{aligned}
\Phi^{(e)}_{I}(x_1,\dots,x_{N_e})&=\frac{1}{\sqrt{N_e!}}\det[\psi_{e,i_a}(x_b)]_{a,b=1}^{N_e},\\
\Phi^{(h)}_{J}(y_1,\dots,y_{N_h})&=\frac{1}{\sqrt{N_h!}}\det[\psi_{h,j_a}(y_b)]_{a,b=1}^{N_h},\\
\ket{D_m^M} &= \Phi^{(e)}_{I}\Phi^{(h)}_{J}
\end{aligned}
\end{equation}
%
with $N_e$ ($N_h$) the number of electrons (holes) in the complex $M$ (e.g., $N_e = 2$, $N_h=1$ for the negative trion X$^-$). Due to spin orbit coupling the orbital and spin parts of $\psi$ cannot be separated, it is, thus, advantageous to write the multi-particle states considered in this work in compact form of second quantization. The multi-particle states are the neutral exciton X
\begin{equation}
\label{eq:suppl:CIWavefunctionX}
\bigl|X\bigr\rangle
=\sum^{n_e}_{i}\sum^{n_h}_{j} \eta^{X}_{ij}\;\hat c_i^\dagger\,\hat d_j^\dagger\,\bigl|\mathrm{GS}\bigr\rangle
\end{equation}
positive trion X$^+$
\begin{equation}
\label{eq:suppl:CIWavefunctionX+}
\bigl|X^+\bigr\rangle
=\sum^{n_e}_{i}\sum^{n_h}_{k<l} \eta^{X^+}_{i;k l}\;
\hat c_i^\dagger\,\hat d_k^\dagger\,\hat d_l^\dagger\,\bigl|\mathrm{GS}\bigr\rangle
\end{equation}
negative trion X$^-$
\begin{equation}
\label{eq:suppl:CIWavefunctionX-}
\bigl|X^-\bigr\rangle
=\sum^{n_e}_{i<j}\sum^{n_h}_{k} \eta^{X^-}_{ij;k}\;
\hat c_i^\dagger\,\hat c_j^\dagger\,\hat d_k^\dagger\,\bigl|\mathrm{GS}\bigr\rangle
\end{equation}
and the neutral biexciton XX
\begin{equation}
\label{eq:suppl:CIWavefunctionXX}
\bigl|XX\bigr\rangle
=\sum^{n_e}_{i<j}\sum^{n_h}_{k<l} \eta^{XX}_{ij;kl}\;
\hat c_i^\dagger\,\hat c_j^\dagger\,\hat d_k^\dagger\,\hat d_l^\dagger\,\bigl|\mathrm{GS}\bigr\rangle
\end{equation}
where $n_e$ and $n_h$ mark the number of single-particle states for electrons and holes in the CI basis, respectively. Moreover, $\hat c_i^\dagger$ creates an electron in conduction spinor orbital $i$, $\hat d_j^\dagger$ creates a hole in valence orbital $j$, and $|\mathrm{GS}\rangle$ marks the fully occupied valence band with electrons. The coefficients $\eta_{m}$ are normalized, i.e. $\sum_m |\eta_{m}|^2=1$. 

Nevertheless, for numerical computational reasons, we still work in our algorithm using Eq.~\eqref{eq:SDgeneralForm} guarding the correct symmetries. Using the aforementioned $\left|D_m^{\rm M}\right>$ the multi-particle trial wavefunction reads
\begin{equation}
    \Psi_i^{\rm M}(\mathbf{r}) = \sum_{\mathit m=1}^{n_{\rm SD}} \mathit \eta_{i,m} \left|D_m^{\rm M}\right>, \label{eq:CIwfSD}
\end{equation}
where $n_{\rm SD}$ is the number of Slater determinants $\left|D_m^{\rm M}\right>$, and $\eta_{i,m}$ is the $i$-th CI coefficient which is found along with the eigenenergy using the variational method by solving the Schr\"{o}dinger equation 
\begin{equation}
\label{CISchrEq}
\hat{H}^{\rm{M}} \Psi_i^{\rm M}(\mathbf{r}) = E_i^{\rm{M}} \Psi_i^{\rm M}(\mathbf{r}),
\end{equation}
where $E_i^{\rm{M}}$ is the $i$-th eigenenergy of the multi-particle state $\Psi_i^{\rm M}(\mathbf{r})$, and~$\hat{H}^{\rm{M}}$ is the CI Hamiltonian which reads
\begin{equation}
\label{eq:CIHamiltonian}
\hat{H}^{\rm{M}}_{mn}=\delta_{mn}\left(\mathcal{E}_m^{{\rm M}(e)}-\mathcal{E}_m^{{\rm M}(h)}\right)+\left<D_m^{\rm M}\right| \hat{V}^{\rm{M}} \left|D_n^{\rm M}\right>,
\end{equation}
where $\delta_{mn}$ is the Kronecker delta and $\mathcal{E}_m^{{\rm M}(e)}$ $\left\{\mathcal{E}_m^{{\rm M}(h)}\right\}$ stands for sum of all single-particle electron (hole) eigenvalues corresponding to eigenstates contained in $\left|D_n^{\rm M}\right>$ for complex $M$. Furthermore, $\left<D_m^{\rm M}\right| \hat{V}^{\rm{M}} \left|D_n^{\rm M}\right>=\sum_{ijkl}V^{\rm{M}}_{ij,kl}$ for $i,j\in S_m$ and $k,l\in S_n$. The sets $S_m$ and $S_n$ contain indices of all single-particle wavefunctions in SDs $\left|D_m^{\rm M}\right>$ and $\left|D_n^{\rm M}\right>$, respectively. Furthermore, $V^{\rm{M}}_{ij,kl}$ is defined by
\begin{equation}
\label{eq:CoulombMatrElem}
\begin{split}
&V^{\rm{M}}_{ij,kl}\equiv(1-\delta_{ij})(1-\delta_{kl})\,q_iq_j\frac{e^2}{4\pi\varepsilon_0}\iint\left(\frac{{\rm d}{\bf r}_1{\rm d}{\bf r}_2}{\epsilon_r(\mathbf{r}_1,\mathbf{r}_2)|{\bf r}_1-{\bf r}_2|}\right)\\
&\times\left(\psi^*_i({\bf r}_1)\psi^*_j({\bf r}_2)\psi_k({\bf r}_1)\psi_l({\bf r}_2)
-\psi^*_i({\bf r}_1)\psi^*_j({\bf r}_2)\psi_l({\bf r}_1)\psi_k({\bf r}_2)\right)\\
&=(1-\delta_{ij})(1-\delta_{kl})\,q_iq_j\left(J^{\rm M}_{ij,kl} - K^{\rm M}_{ij,lk}\right),
\end{split}
\end{equation}
where $\varepsilon_0$ and $\epsilon_r(\mathbf{r}_1,\mathbf{r}_2)$ are the vacuum and spatially dependent relative permittivity, respectively, and $\delta_{ij}$ and $\delta_{kl}$ are the Kronecker deltas. Note that the terms in the first two brackets in Eq.~\eqref{eq:CoulombMatrElem} ensure that each single-particle state in SD occurs only once, thus preventing double counting. Furthermore, $q_i,q_j\in\{-1,1\}$ marks the sign of the charge of the quasiparticles in states with indices $i$ and $j$, respectively; $e$ is the elementary charge. The parameters $J^{\rm M}$ and $K^{\rm M}$ in Eq.~\eqref{eq:CoulombMatrElem} are direct and exchange Coulomb integrals. 

Since the single-particle states are orthonormal, one finds that in Eq.~\eqref{eq:CIHamiltonian} there are only three possible kinds of matrix elements in CI,~i.e.

\begin{widetext}
\begin{equation}  
\label{eq:CIHamiltonianSeparated}
\begin{split}
\hat{H}^M_{mn} &=   \begin{cases}
    \mathcal{E}_m^{{\rm M}(e)}-\mathcal{E}_m^{{\rm M}(h)} 
    + \dfrac{1}{2}\sum\limits_{i,j\in S_n} &\left(J^{\rm M}_{ij,ij} - K^{\rm M}_{ij,ji}\right)
     \text{  if $m = n$}\\
      \dfrac{1}{2} \sum\limits_{j\in S_n} \left(J^{\rm M}_{ij,kj} - K^{\rm M}_{ij,jk}\right) & \text{if $D^M_m$ and $D^M_n$ differ by one single-particle state: $\ket{D^M_m} \propto c^\dagger_i c_k \ket{D^M_n}$ } \\
      \dfrac{1}{2} \left(J^{\rm M}_{ij,kl} - K^{\rm M}_{ij,lk}\right) & \text{if $D^M_m$ and $D^M_n$ differ by two single-particle states: $\ket{D^M_m} \propto c^\dagger_i c^\dagger_j  c_k c_l \ket{D^M_n}$ , $k<l$}.
    \end{cases}
  \end{split}
\end{equation}  
\end{widetext}

\subsection{Method of calculation of configuration interaction}
\label{subsec:CIwaycalc}
The sixfold integral in Eq.~\eqref{eq:CoulombMatrElem} is evaluated using the~Green's function method~\cite{Schliwa:09,Klenovsky2017}. The integral in Eq.~\eqref{eq:CoulombMatrElem} is divided into a solution of Poisson's equation for one quasiparticle $a$ only, followed by a three-fold integral for the quasiparticle $b$ in the electrostatic potential generated by the particle $a$ and resulting from the previous step. That procedure, thus, makes the whole solution numerically more feasible and is described by
\begin{equation}
\begin{split}
\nabla\!\cdot\!\big[\varepsilon_0\,\varepsilon_r(\mathbf r_1)\,\nabla \hat{U}_{ajl}(\mathbf r_1)\big]
= -\,q_a e\, \Psi^*_{aj}(\mathbf r_1)\Psi_{al}(\mathbf r_1),\\
V^{M}_{ij,kl} = \int d^3 r_2\, \hat{U}_{ajl}(\mathbf r_2)\,(q_b e)\,\Psi^*_{bi}(\mathbf r_2)\Psi_{bk}(\mathbf r_2)\,.
\end{split}
\label{eq:GreenPoisson}
\end{equation}
where $a,b \in \{e,h\}$ and we have assumed that the spatial vectors $\mathbf{r}_1$ and $\mathbf{r}_2$ span the same space.

\subsection{Radiative rate \& lifetime}
\label{subsec:radiate}
Following Stobbe~{\sl et~al.} (see Ref.~\cite{Stobbe2012}, Eq.~(21) and App.~C), the spontaneous-emission rate of a many-body state $|i\rangle$
can be written as
\begin{equation}
\label{eq:Stobbe_master}
\Gamma_i(\omega)=\frac{2}{\hbar}\iint \mathbf J_i^*(\mathbf r)\cdot
\mathrm{Im}\,\mathbf G(\mathbf r,\mathbf r';\omega)\cdot \mathbf J_i(\mathbf r')\,d^3\mathbf r\,d^3\mathbf r',
\end{equation}
where the interband transition current is
\begin{equation}
\label{eq:Ji_CI_decomp}
\mathbf J_i(\mathbf r)=
\sum_{m=1}^{n_{\rm SD}} \eta_{i,m}
\sum_{(r,q)\in D_m^{\rm M}} \mathbf J^{(rq)}(\mathbf r),
\end{equation}
\begin{equation}
\label{eq:Jrq_env}
J^{(rq)}_{\alpha}(\mathbf r)=\frac{e}{m_0}
\sum_{\nu_v\in V}\sum_{\nu_c\in C}
\chi_{h_r,\nu_v}^{*}(\mathbf r)\,p_{\alpha,\nu_v\nu_c}\,\chi_{e_q,\nu_c}(\mathbf r),
\end{equation}
where $\alpha\in\{x,y,z\}$. In a homogeneous background Eq.~\eqref{eq:Stobbe_master} factorizes into a material local density of states (LDOS) prefactor and a transition amplitude,
\begin{equation}
\label{eq:GammaClasOsc}
\begin{aligned}
&\Gamma_{i,\mu}^{\rm M}(E_i)=\Gamma_{\rm cl}(E_i)\,f^{\rm M}_{i,\mu},\\
&\Gamma_{\rm cl}(E)=\frac{n(E)\,e^2 E^2}{6\pi m_0\varepsilon_0\hbar^2 c^3},\\
&E_i=\hbar\omega_i,
\end{aligned}
\end{equation}
where $\mu$ denotes the detected polarization and $n(E)$ is the dispersive refractive index. In Eq.~\eqref{eq:GammaClasOsc} we define $\Gamma_{\rm cl}(E)$ for a \emph{single} linear polarization. 
The total radiative rate for transition $i$ is obtained by summing over the two transverse polarizations,
\begin{equation}
\label{eq:GammaSum}
\Gamma^{\rm M}_i(E) = \Gamma_{\rm cl}(E)\sum_{\mu \in \{x,y\}} f^{\rm M}_{i,\mu}\, .
\end{equation}

\emph{Dipole approximation (DA).} Approximating the extended current $\mathbf J$ by a point dipole yields the standard DA oscillator strength.
At the envelope-function level (spinor indices $\nu_v\in V$ for the valence block and $\nu_c\in C$ for the conduction block) we obtain
\begin{equation}
\label{eq:CIOscStrengthPol}
\begin{aligned}
f^{\rm M}_{i,\mu,{\rm DA}}
&=\frac{2}{m_0 E_i}\left|
\sum_{m=1}^{n_{\rm SD}}\eta_{i,m}
\sum_{(r,q)\in D_m^{\rm M}}
\sum_{\nu_v\in V}\sum_{\nu_c\in C}\times\right.\\
&\left.\times\int d^3\mathbf r\;
\chi_{h_r,\nu_v}^{*}(\mathbf r)\,
\big(\hat{\mathbf e}_\mu\!\cdot\!\mathbf p_{\nu_v\nu_c}\big)\,
\chi_{e_q,\nu_c}(\mathbf r)
\right|^2 ,
\end{aligned}
\end{equation}
with $\mathbf p_{\nu_v\nu_c}=\langle u^\Gamma_{\nu_v}|\hat{\mathbf p}|u^\Gamma_{\nu_c}\rangle$ (Kane $p$-form; $r/p$ gauge equivalence holds within the 8-band model).

\emph{Beyond-dipole approximation (BDA).} Retaining the finite emitter size corresponds to keeping the longitudinal projection of the current in Eq.~\eqref{eq:Stobbe_master}, yielding the BDA oscillator strength 
\begin{equation}
\label{eq:CI_BDA_final}
\begin{aligned}
&f^{\rm M}_{i,\mu,{\rm BDA}}
=\frac{2m_0}{e^2 E_i}\Bigg|
\sum_{m=1}^{n_{\rm SD}}\eta_{i,m}
\sum_{(r,q)\in D_m^{\rm M}}
\iint d^3\mathbf r\,d^3\mathbf r'\times\\
&\times\,
\hat{\mathbf e}_\mu\!\cdot\!
\Big[\nabla_{\!r}\nabla_{\!r'}\,\frac{1}{4\pi|\mathbf r-\mathbf r'|}\Big]\!\cdot\!
\mathbf J^{(rq)}(\mathbf r')\Bigg|^2 .
\end{aligned}
\end{equation}
%
%
If the longitudinal projection is omitted, Eq.~\eqref{eq:CI_BDA_final} reduces to the dipole-approximation result, which is equivalent to Eq.~\eqref{eq:CIOscStrengthPol} upon using Eq.~\eqref{eq:Jrq_env}.
All beyond-dipole effects enter via the longitudinal projection acting on the extended current, while the LDOS prefactor remains homogeneous (transverse Im\,$\mathbf G_T$). Note that Eq.~\eqref{eq:CI_BDA_final} is written for a homogeneous background permittivity. For spatially varying $\varepsilon_r(\mathbf r)$ we employ the equivalent Poisson formulation~\eqref{eq:CIOscStrengthPoissonWK}–\eqref{eq:CIOscStrengthPolWK} which we discuss in the following.

An equivalent Poisson form is obtained by introducing
\begin{equation}
\label{eq:OscStrPoissonSource}
\rho_{\rm eff}^{(rq)}(\mathbf r)=\frac{1}{i\omega_{rq}}\nabla\!\cdot\!\mathbf J^{(rq)}(\mathbf r),
\end{equation}
where $\omega_{rq}\equiv \omega_i$ and solving
\begin{equation}
\label{eq:CIOscStrengthPoissonWK}
\nabla\!\cdot[\varepsilon_0\varepsilon_r(\mathbf r)\nabla\Phi_{rq}(\mathbf r)]=-\,\rho_{\rm eff}^{(rq)}(\mathbf r),
\end{equation}
yielding
\begin{equation}
\label{eq:CIOscStrengthPolWK}
\begin{aligned}
f^{\rm M}_{i,\mu,{\rm BDA}}&=\frac{2m_0}{e^2 E_i}\Bigg|\sum_{m,(r,q)}\eta_{i,m}\,\times\\
&\times\int d^3\mathbf r\,\hat{\mathbf e}_\mu\!\cdot\!\left[-i\omega_i\varepsilon_0\varepsilon_r(\mathbf r)\nabla\Phi_{rq}(\mathbf r)\right]\Bigg|^2.
\end{aligned}
\end{equation}
%
%
%
Note that in Eqs.~\eqref{eq:CI_BDA_final} and~\eqref{eq:CIOscStrengthPolWK}, both the electron and hole spinors enter only through the transition current 
$\mathbf J^{(rq)}$ defined in Eq.~\eqref{eq:Jrq_env}, i.e., they are treated on equal footing.
The equations~\eqref{eq:CIOscStrengthPoissonWK}~and~\eqref{eq:CIOscStrengthPolWK} are solved in this work for BDA, while Eq.~\eqref{eq:CIOscStrengthPol} is solved in case of DA. The radiative lifetime is then computed from Eq.~\eqref{eq:GammaSum} as
\begin{equation}
\label{eq:lifetimeFromGamma}
\tau^{\rm M}_{i}=1/\Gamma^{\rm M}_{i}(E_i).    
\end{equation}

\section{Results}
\label{sec:results}
\subsection{Exciton in GaAs/AlGaAs QDs}
\label{subsec:exciton}
\begin{figure}[htbp]
    \includegraphics[width=85mm]{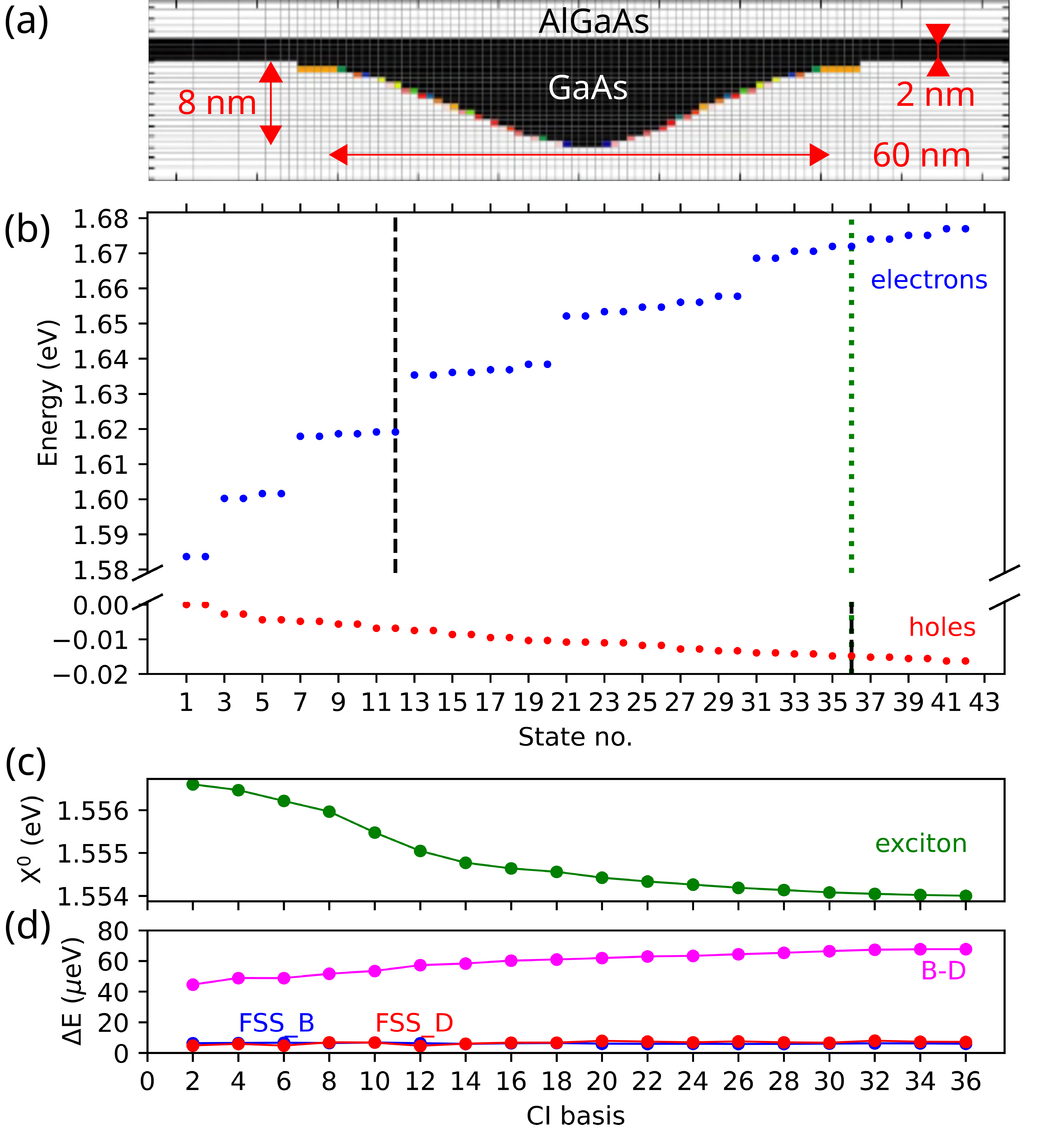}	
	\caption{The simulated structure of GaAs ``QD1" with 2~nm GaAs wetting layer (WL) in Al$_{0.4}$Ga$_{0.6}$As is shown in panel (a) with marked QD and WL dimensions~\cite{Yuan2023,yuan_xueyong_2023_7748664}. Panel (b) gives the single-particle energies of the simulated QD for electrons (blue symbols) and holes (red symbols). For each kind of quasiparticle the energies of 42 states are shown in (b). The doubling of states for each energy level in (b) corresponds to the Kramers doublets of corresponding states. The black broken and green dotted vertical lines in (b) correspond to the largest CI bases used in this work for computations of ${\rm M}\in\{{\rm X}^-, {\rm X}^+, {\rm XX}\}$ and that for X$^0$, respectively. In panel (c) the ground state exciton energy (X$^0$) is shown (by green balls) as a function of symmetric CI basis size. The exciton energy reaches a value of X$^0$=1.5541~eV for a CI basis of 36 $\psi^{(e)}$ and 36 $\psi^{(h)}$ (36x36 CI basis). For comparison, the measured value of X$^0$ was~1.551152~eV~\cite{Yuan2023}. Panel (d) shows the evolution of bright (FSS\_B) and dark (FSS\_D) FSS of X$^0$ in blue and red balls, respectively, on symmetric CI basis size. That for the bright-dark (B-D) splitting of X$^0$ is given in (d) by violet balls. We see that computed bright FSS value of $7\pm0.5\,\mu$eV almost does not change with size of CI basis while B-D splitting ceases to change appreciably when reaching a value of 68~$\mu$eV. Note that a more detailed analysis of convergence of energies of X$^0$ and B-D splitting in panels (c) and (d) is given in Fig.~\ref{fig:Econv}~(a)~and~(b) in the Appendix~I.
    }
	\label{fig:AFMsp}
\end{figure} 

{\color{black}
In this work, we consider realistic GaAs/Al$_{0.4}$Ga$_{0.6}$As QD defined using AFM nanohole scan in Fig.~\ref{fig:AFMsp}~(a), being the same as ``QD1" in Refs.~\cite{Yuan2023,yuan_xueyong_2023_7748664}.
}
In Fig.~\ref{fig:AFMsp}~(b) 42 single-particle energies of electrons and holes for QD defined in (a) are given by blue and red balls, respectively. The computed energies of holes are much more closely spaced than those of electrons~\cite{Schliwa:09}. That is a consequence of the different effective masses being $0.067\,m_0$ and $0.51\,m_0$ for electrons and heavy holes in GaAs~\cite{Vurgaftman2001}, respectively. 

In Fig.~\ref{fig:AFMsp}~(c) the evolution of the ground state exciton (X$^0$) energy with symmetric CI basis (i.e.~the same number of $\psi^{(e)}$ and $\psi^{(h)}$) is shown.
The decrease of X$^0$ energy change with nominal increase of CI basis size is observed (see Fig.~\ref{fig:Econv}~(a) in Appendix~I.). For CI basis of 36 $\psi^{(e)}$ and 36 $\psi^{(h)}$ that change is less than 8~$\mu$eV and a value of X$^0$ energy of 1.5541~eV is found. That value is larger by only 3~meV than the experimentally observed value of X$^0=1.551152$~eV~\cite{Yuan2023}.

Furthermore, in Fig.~\ref{fig:AFMsp}~(d) the CI basis convergence study is given also for bright and dark X$^0$ fine-structure splitting (FSS) by blue and red balls, respectively. Both quantities show negligible dependence on CI basis size, maintaining values of $7\pm0.5\,\mu$eV and $6\pm0.5\,\mu$eV for bright and dark FSS, respectively. Note that the experimental value of bright FSS was measured as 8.1~$\mu$eV~\cite{Yuan2023}. Moreover, in Fig.~\ref{fig:AFMsp}~(d) the variation of the energy separation between bright and dark X$^0$ doublet (B-D) is shown by violet balls. That energy separation increases with CI basis size, reaching a value of $\approx 68\,\mu$eV for 36x36 CI basis. At that point the nominal change in B-D splitting energy with CI basis increase is less than $0.05\,\mu$eV (see also Fig.~\ref{fig:Econv}~(b) in Appendix~I.). Sadly, the calculated value of B-D splitting does not reach the experimental value of $\approx 100\,\mu$eV~\cite{Yuan2023}. Nevertheless, taken together we can still conclude that the $\mathbf{k}\!\cdot\!\mathbf{p}$~+~CI calculations very well reproduce the experimental results on exciton published elsewhere~\cite{Yuan2023}.

\subsection{Multi-particle complexes in GaAs/AlGaAs QDs}
\label{subsec:multiparticle}
We now turn our attention to multi-particle complexes. For complexes consisting of more than one electron or one hole, the key numerical issue in CI implementation is related to the combinatorial complexity of generating all available SDs for a given number of single-particle CI basis states~\cite{Shumway2001,Rontani2006,Troparevsky2008,Schliwa:09}. {\color{black} Note that in the case when both the number of interacting fermions as well as the number of CI basis states are increased at the same time, the number of SDs grows approximately {\sl exponentially} when Stirling's formula is employed~\cite{Sherrill1999}.
However, the convergence of CI is studied in this work by increasing the number of single-particle CI basis states, leading to a {\sl polynomial} growth~\cite{Sherrill1999,Shumway2001} of the number of SDs. To reduce the CI space, we employ here singles–doubles CI (SDCI),~i.e., we include all SDs that differ from a reference occupation by at most one or two spin-orbital substitutions (single and double excitations)~\cite{Sherrill1999,brasken2000full,Schliwa:09,Purvis1982}}.
Another possibility of reducing the number of SDs is to consider an asymmetric CI basis,~i.e., with different numbers of $\psi^{(e)}$ and $\psi^{(h)}$. That is verified by the fact that the energy densities of $\mathcal{E}^{(e)}$ and $\mathcal{E}^{(h)}$ are markedly different, see Fig.~\ref{fig:AFMsp}~(b). Note that in Fig.~\ref{fig:AFMsp}~(b) all computed $\mathcal{E}^{(e)}$ span 93~meV, while the same number of $\mathcal{E}^{(h)}$ spans only 16~meV.

\begin{figure*}[htbp]
    \centering
    \includegraphics[width=170mm]{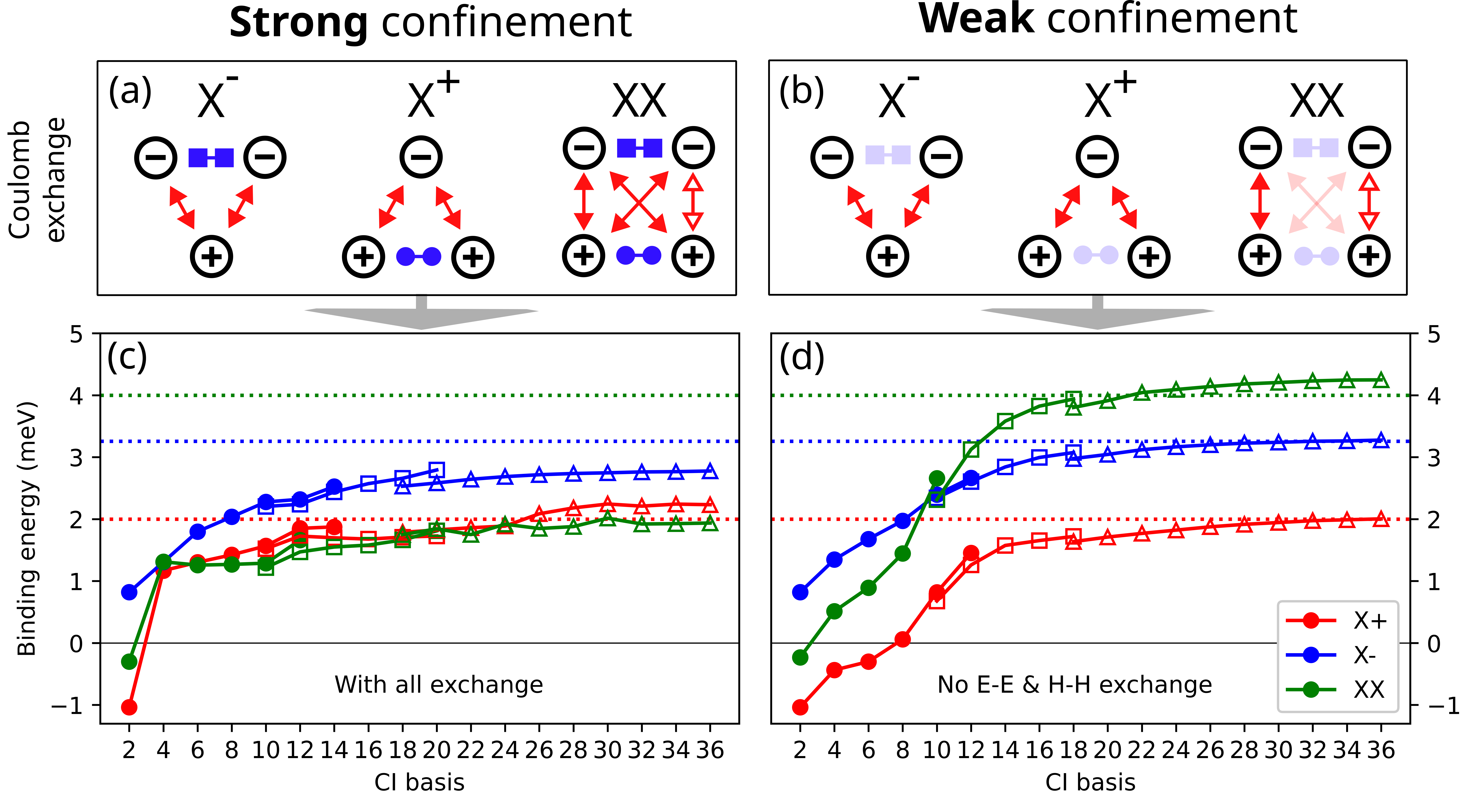}	
	\caption{Panels (a) and (b) show the sketches of the type of the Coulomb exchange considered in CI calculations for X$^-$, X$^+$, and XX. In (a) and (b) the red triangles mark the electron-hole ($K_{eh}$), blue boxes the electron-electron ($K_{ee}$), and balls hole-hole ($K_{hh}$) Coulomb exchange interaction. The empty symbols in (a) and (b) for XX mark $K_{eh}$ of one of the final states of the recombination of XX, i.e. X$^0$. The dimmed colored lines and symbols in (b) mark the exchange interactions omitted in the CI calculations in (d) (see text). In (c) and (d) the variations with respect to the number of single-particle states in CI basis for the binding energies of X$^-$, X$^+$, and XX relative to X$^0$ are shown. In correspondence to (a) and (b), in (c) all Coulomb direct and exchange integrals are considered, while in (d) $K_{ee}$ and $K_{hh}$, and partly $K_{eh}$ are omitted. The meaning of markers in (c) and (d) is the following: (i) full balls represent symmetric CI basis,~i.e., same number of $\psi^{(e)}$ and $\psi^{(h)}$; (ii) open squares represent the same but for SDCI approximation; (iii) open upward triangles give SDCI for asymmetric basis composed of twelve $\psi^{(e)}$ and varying number of $\psi^{(h)}$.
    Note that there is a negligible energy offset $<100\,\mu$eV between the calculations performed using aforementioned methods, seen as steps for the overlapping CI bases \{e.g. CI bases of 10 and 18 in (c) and (d)\}.
    The red horizontal broken line denotes experimental binding energy of X$^+$~\cite{Yuan2023}, blue of X$^-$~\cite{Huber2019}, and green of XX~\cite{DaSilva2021}. Notice that calculations reach very close to experimental values of binding energies in (d), i.e., for calculation with $K_{ee}$, $K_{hh}$ and partly $K_{eh}$ omitted [dimmed colored arrows in (b)], corresponding to the situation due to weak confinement effect, see also main text. Note that a more detailed analysis of convergence of binding energies of X$^+$, X$^-$ and XX is given in Fig.~\ref{fig:Econv}~(c) in the Appendix~I.
    }
	\label{fig:BindingEXnoEEHH}
\end{figure*}
In Fig.~\ref{fig:BindingEXnoEEHH} the evolution of binding energies of X$^-$, X$^+$, and XX with respect to X$^0$ with the number of CI basis states is shown. Due to the numerical complexity of the CI previously discussed, three levels of approximations are used with an increase of the CI basis size: (i) symmetric CI basis,~i.e., same number of $\psi^{(e)}$ and $\psi^{(h)}$; (ii) the same as for the previous point but for SDCI approximation; (iii) SDCI for the asymmetric CI basis composed of twelve $\psi^{(e)}$ and variable number of $\psi^{(h)}$. In all CI and SDCI calculations of the complexes in this work, the direct Coulomb integrals ($J$) between all quasiparticles are considered. However, two scenarios are discussed for the Coulomb exchange interaction ($K$) as indicated in Fig.~\ref{fig:BindingEXnoEEHH}~(a)~and~(b). In~(a), all Coulomb exchange is considered between all quasiparticles, while in~(b) the electron-electron ($K_{ee}$), hole-hole ($K_{hh}$) and part of the electron-hole ($K_{eh}$) Coulomb exchange interactions are neglected [neglected exchange interactions are marked by dimmed colored lines and arrows in~(b)].

In agreement with previous reports~\cite{Schliwa:09}, in Fig.~\ref{fig:BindingEXnoEEHH}~(c)~and~(d) without correlation X$^-$ is found to be binding while X$^+$ and XX are anti-binding. An increase in the size of the CI basis and associated correlation causes X$^+$ and XX to also become binding. The smallest increase in the binding energies of X$^-$, X$^+$, and XX is reached in Fig.~\ref{fig:BindingEXnoEEHH}~(c)~and~(d) for the SDCI with the basis consisting of 12 $\psi^{(e)}$ and 36 $\psi^{(h)}$ which is called the 12x36 SDCI basis in the following~\cite{supkptest}. Although in Fig.~\ref{fig:BindingEXnoEEHH}~(c) binding energy of X$^+$ increases towards the experimental value~\cite{Yuan2023}, that for X$^-$ reaches a magnitude somewhat smaller than reported in the measurements~\cite{Huber2019}. However, the calculations preserve at least the binding energy ordering of X$^+$ and X$^-$,~i.e. the magnitude of the former (X$^+$) being smaller. Sadly, calculations for binding energy of XX miss the experimental target by almost 2~meV. Note that a similar disagreement with experimental results as in Fig.~\ref{fig:BindingEXnoEEHH}~(c) was previously observed for smaller GaAs QDs~\cite{Wang2009}.

The convergence towards the experiment for the 12x36 SDCI basis is considerably improved for all complexes in Fig.~\ref{fig:BindingEXnoEEHH}~(d), where $K_{ee}$ and $K_{hh}$ and partly $K_{eh}$ are neglected~\cite{Honig2014}. The improvement is particularly striking for XX, the binding energy of which almost doubles between Fig.~\ref{fig:BindingEXnoEEHH}~(c)~and~(d) reaching very close to the measured value. Similarly as for X$^0$ in Fig.~\ref{fig:AFMsp} the convergences of the computed binding energies of X$^-$, X$^+$, and XX relative to X$^0$ are shown in more detail in Fig.~\ref{fig:Econv}~(c). There we can see that $|\Delta E/\Delta N|$, where $\Delta E$ marks the difference in binding energies for two consecutive CI basis state sizes and $N$ marks the number of CI basis states, is $<10\,\mu$eV for the 12x36 CI basis,~i.e. two orders of magnitude smaller than the absolute values of the binding energies for all studies complexes. We note that in all calculations for the 12x36 basis the energy of the correlated electron-hole exchange interaction $K_{eh}$ is 0.01~meV for trions and 0.18~meV for biexciton confirming that correlated direct Coulomb interaction $J$ mainly causes the large binding energy of complexes in Fig.~\ref{fig:BindingEXnoEEHH}~(c)~and~(d)~\cite{Schliwa:09}. The difference between Fig.~\ref{fig:BindingEXnoEEHH}~(c)~and~(d) is solely in the amount of correlated $K_{ee}$ and $K_{hh}$ (and partly $K_{eh}$, which however cannot account for the difference~\cite{supehexchange}), which naturally also depend on the complex $M$. Furthermore, note that by comparing the binding energies in Fig.~\ref{fig:BindingEXnoEEHH}~(c)~and~(d) it follows that the effect of correlated $K_{ee}$ and $K_{hh}$ on X$^-$ and XX is anti-binding, similar to that for the direct Coulomb interaction $J_{ee}$.

The suppression of exchange interactions $K_{ee}$ and $K_{hh}$ (and partly $K_{eh}$) in multi-excitonic complexes in GaAs/AlGaAs QDs can be understood by considering the asymmetry between direct and exchange Coulomb interactions, especially in large systems exhibiting weak confinement effects~\cite{Huber2019,Alshaikh2024}. For a QD with a base diameter of 60~nm and a height of 8~nm, as discussed in this work, the spatial extent of the electron and hole wavefunctions becomes comparable to or exceeds their correlation length. In such systems, the overlap of the fermionic orbitals becomes very sensitive to the distance between quasiparticles. Thus, even a rather small spatial separation due to direct Coulomb repulsion $J_{ee}$ or $J_{hh}$ between quasiparticles carrying the same charge in X$^-$, X$^+$ and XX might lead to severe suppression of the exchange integrals $K_{ee}$ and $K_{hh}$ which generally scale as $\sim1/r^3$~\cite{Takagahara2000,Huo2014,Krapek2015} where $r$ is the distance between quasiparticles. In contrast, the direct Coulomb interaction $J_{ee}$ (or $J_{hh}$) reduces with $r$ as $\sim 1/r$ and remains substantial because it depends primarily on the charge distribution and not on the overlap of the wavefunctions.

The aforementioned situation can naturally arise,~e.g., under quasi-resonant excitation conditions, where specific many-body states are selectively populated. For example, configurations with delocalized electron orbitals (due to their lower effective mass) but strongly confined holes may exhibit suppressed $K_{ee}$ and finite $K_{hh}$. Similarly, the mixed-spinor structure from spin–orbit coupling can suppress hole exchange in specific symmetry-adapted configurations. A key feature in the case of resonant excitation is the {\sl separation of charges}, which enhances the emission of particular complexes.

\subsection{Radiative lifetime of GaAs/AlGaAs QDs}
\begin{figure*}[htbp]
    \includegraphics[width=170mm]{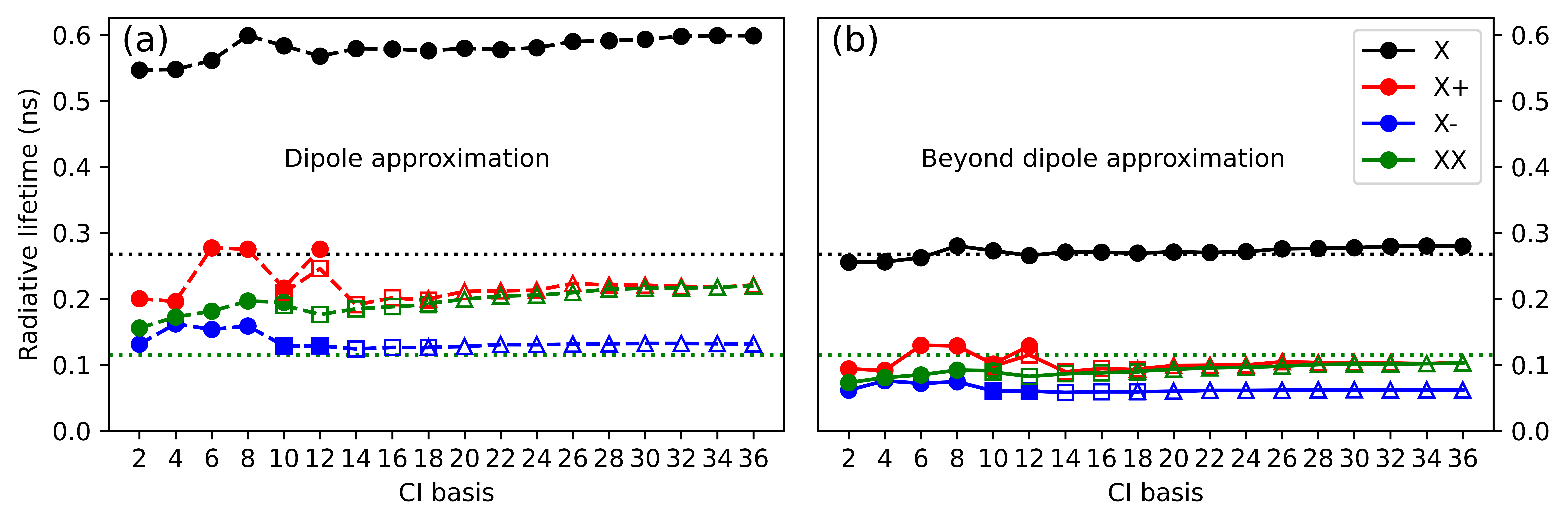}
	\caption{The evolution of the radiative lifetime of ground states of X, X$^+$, X$^-$, and XX as a function of the CI basis size when the overlap integrals are evaluated considering (a)~DA and (b) BDA~\cite{Stobbe2012}, see also main text. The meaning of markers in both panels is the same as that in Fig.~\ref{fig:BindingEXnoEEHH}~(c)~and~(d). The black (green) dotted horizontal line marks the measured values of exciton (biexciton) lifetime from Ref.~\cite{Schimpf2019}. Note that for both DA and BDA the calculations of lifetime do not change appreciably for CI bases larger than $14$ states. On the other hand, the calculations using BDA reproduce the experiments considerably better than those for DA.}
	\label{fig:Lifetime}
\end{figure*} 
We now discuss the calculations of radiative lifetime $\tau^{\rm M}$ of the complexes $M$ discussed in the previous section. The evolution of $\tau^{\rm M}$ with CI basis size for $M\in\left\{{\rm X}^0,\,{\rm X}^+,\,{\rm X}^-,\,{\rm XX}\right\}$ is shown in Fig.~\ref{fig:Lifetime}~(a)~and~(b) for the case of DA and BDA, respectively, see also Eqs.~\eqref{eq:CIOscStrengthPol}~and~~\eqref{eq:CIOscStrengthPolWK} in Sec.~\ref{subsec:radiate}. The multi-particle calculation for X$^+$, X$^-$, and XX in Fig.~\ref{fig:Lifetime} are performed with omitted exchange integrals $K_{ee}$, $K_{hh}$ and partly $K_{eh}$ similarly as in Fig.~\ref{fig:BindingEXnoEEHH}~(d)~\cite{Honig2014}.

Firstly, one can see in Fig.~\ref{fig:Lifetime} that lifetime $\tau^{\rm M}$ for all studied complexes converges already for a rather small ($<14$) CI basis size. Even the smallest CI basis of two electron and two hole single-particle states provides a very good estimation of the Coulomb correlated emission lifetime of complexes. 

Secondly, in Fig.~\ref{fig:Lifetime}~(b) we see that for the case of BDA the lifetimes of X$^0$ and XX converge to values of $\tau^{\rm X}=0.279$~ns and $\tau^{\rm XX}=0.101$~ns, while for just DA in Fig.~\ref{fig:Lifetime}~(a), the corresponding values are $0.598$~ns and $0.217$~ns, respectively. The reported experimental values of X$^0$ and XX lifetimes are $0.267$~ns and $0.115$~ns~\cite{Schimpf2019}, respectively, and are marked by black and green broken horizontal lines in Fig.~\ref{fig:Lifetime}. Clearly, the computed results obtained for BDA are much closer to the experimental values than those for DA.

The ratio of the computed lifetimes in panel (b) relative to panel (a) of Fig.~\ref{fig:Lifetime} is $\approx 0.47$ and is approximately similar for all computed complexes. That lifetime reduction is connected with the size of the QD body and is associated with the weak confinement regime~\cite{Stobbe2010,Stobbe2012}. To confirm that, we have studied the size dependence of X$^0$ lifetime for another GaAs/Al$_0.4$Ga$_0.6$As QD and show the results in Fig.~\ref{fig:LifeVdep}~(b) in Appendix II. We can clearly see from that figure that while results for DA do not depend on QD size appreciably (except for the largest dots), the lifetime of X$^0$ computed using BDA progressively reduces with QD size. For consistency reasons, we discuss in Fig.~\ref{fig:LifeVdep}~(a) of Appendix II. also the QD size dependence of FSS and the B-D splitting of X$^0$, the latter showing considerable dependence on QD volume which might be one of the possible reasons for not completely fitting the value of B-D between theory and experiment in Fig.~\ref{fig:AFMsp}~(d).

\subsection{Electric field dependence of GaAs/AlGaAs QDs}
\label{subsec:elfield}
\begin{figure}[htbp]
    \includegraphics[width=90mm]{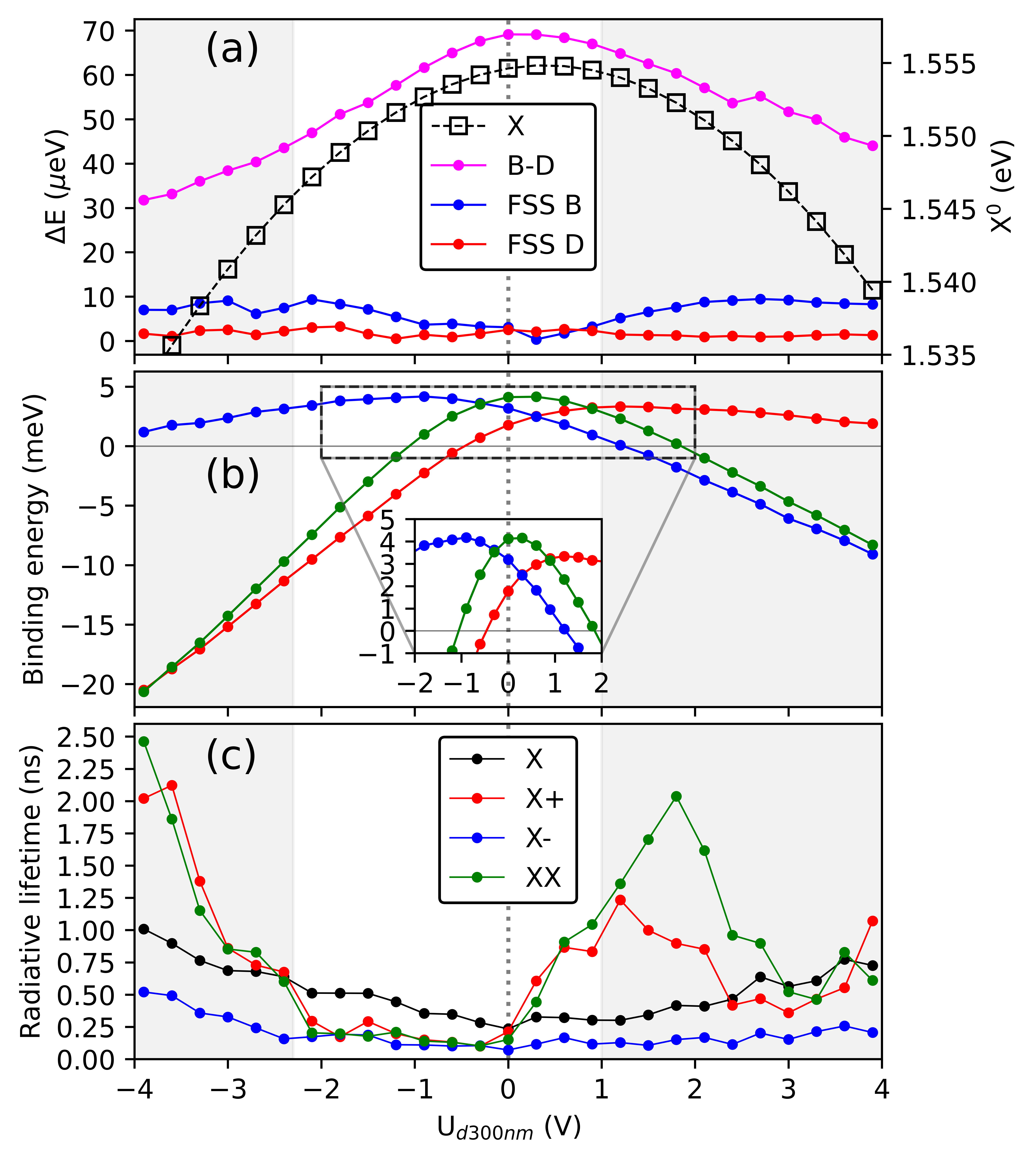}
	\caption{Panel (a) gives the vertical electric field dependence of X$^0$ energy (open squares, values on the right vertical axis), B-D splitting (full violet balls), bright FSS (full blue balls) and dark FSS (full red balls) of X$^0$. The values of latter three parameters are on the left vertical axis. In panel (b) we show the evolution of the binding energy of X$^+$ (red), X$^-$ (blue), and XX (green) relative to X$^0$ with vertically applied electric field on QD in Fig.~\ref{fig:AFMsp}~(a). The inset in (b) shows an enlarged part of the data corresponding to the band crossings. The meaning of axes in the inset are the same as for the whole panel (b). In (c) we give the radiative lifetime of X$^0$, X$^+$, X$^-$, and XX computed using BDA. In order to facilitate the comparison with Ref.~\cite{Undeutsch2025}, the electric field is given as a voltage applied on 300~nm thick layer, hence the label of horizontal axis of $U_{d300nm}$. The data coloring in (c) is the same as in (b) except for X$^0$ which is given in black. The curves in both panels are guides to the eye. The gray-shaded areas in all panels correspond to voltages not considered in Ref.~\cite{Undeutsch2025}. The calculations of X$^0$ were performed with the CI basis of 36 electron and 36 hole single-particle states while that for X$^+$, X$^-$ and XX using SDCI with basis of 12 electron and 36 hole states and with omitted $K_{ee}$, $K_{hh}$ and partly $K_{eh}$ exchange integrals see Fig.~\ref{fig:BindingEXnoEEHH}~(d).}
	\label{fig:ELfldBindLife}
\end{figure}
In order to further test our previously discussed theory, we have computed the evolution of properties of X$^0$, binding energies of X$^+$, X$^-$, XX and the lifetime of those in vertical electric field, see Fig.~\ref{fig:ELfldBindLife}. Our aim was to compare our computed results with experiments discussed by Undeutsch~{\sl et~al.}~in Ref.~\cite{Undeutsch2025}. In our calculations the same GaAs/Al$_{0.4}$Ga$_{0.6}$As QD as that in Fig.~\ref{fig:AFMsp}~(a) was used (different from that studied in Ref.~\cite{Undeutsch2025}), but the vertically applied electric field with the same orientation and magnitudes was considered as in Ref.~\cite{Undeutsch2025}. In the following, we specify the magnitude of the electric field by providing the applied voltage $U_{d300nm}$ on a layer with a thickness of $d=300$~nm. The electric field magnitude is then clearly specified as $U_{d300nm}/d$ and consequently the voltage scale is the same as that used in Ref.~\cite{Undeutsch2025} to ease comparison.

In Fig.~\ref{fig:ELfldBindLife}~(a) the energy structure of X$^0$ in the vertical electric field is shown. We see a clear Stark shift of X$^0$ energy with maximum at 1.5548~eV occurring for the electric field corresponding to applied voltage of $U_{d300nm}=0.3$~V related to an electric field of 10~kV/cm. For the same value of $U_{d300nm}$ we observe in Fig.~\ref{fig:ELfldBindLife}~(a) the maximum B-D splitting of $69\,\mu$eV. Similarly as for X$^0$ energy, the B-D splitting follows the Stark curve and is reduced in magnitude for $U_{d300nm}=\pm4$~V to $30-40\,\mu$eV. The bright FSS first decreases with $U_{d300nm}>0$ to a negligible value of $\approx0.36\,\mu$eV at $U_{d300nm}=0.3$~V, i.e. field of 10~kV/cm, similar to Refs.~\cite{Ghali2012,Luo2012}. The crossing of minimal value of bright FSS is associated in our calculation with rotation of polarization axis of bright X$^0$. Further increase of $U_{d300nm}$ from the bright FSS minimum to positive or negative values results in increase of bright FSS magnitude. On the other hand, dark FSS is affected by electric field far less and has a mean value of $1.7\pm0.5\,\mu$eV.

In Fig.~\ref{fig:ELfldBindLife}~(b) the evolution of binding energy of X$^+$, X$^-$, and XX relative to X$^0$ with $U_{d300nm}$ is shown. The binding energy of XX reduces from its maximum again attained at $U_{d300nm}=0.3$~V with increase towards both positive and negative values of $U_{d300nm}$. Crossings with X$^0$ \{i.e. crossings of zero in Fig.~\ref{fig:ELfldBindLife}~(b)\} are obtained for $-1.1$~V and $1.8$~V, the former being close to experimental value of ca. $-1.5$~V in Ref.~\cite{Undeutsch2025}. The dependence of X$^+$ and X$^-$ binding energies on $U_{d300nm}$ is considerably asymmetric and different to that of XX. For negative values of $U_{d300nm}$ binding energy of X$^-$ first increase up to $4.2$~meV for $U_{d300nm}=-0.9$~V and then slowly decrease. On the other hand, for $U_{d300nm}>0$ the decrease in binding energy of X$^-$ is more rapid and is similar to that for XX. For the binding energy of X$^+$ a reversed scenario is observed. For that the increase of the binding energy occurs for $U_{d300nm}>0$ with maximum of $3.3$~meV attained at $U_{d300nm}=1.2$~V followed by further decrease of binding energy. However, the rapid decrease of X$^+$ binding energy occurs for $U_{d300nm}<0$. The rate of the decrease of binding energy of X$^-$ for $U_{d300nm}>0$ (X$^+$ for $U_{d300nm}<0$) is somewhat smaller than that of the binding energy of XX. That leads to the crossing of X$^-$ and XX (X$^+$ and XX) at $U_{d300nm}=4$~V ($U_{d300nm}=-4$~V).

Furthermore, in Fig.~\ref{fig:ELfldBindLife}~(c) the computed dependence of the radiative lifetime $\tau$ of X$^0$, X$^+$, X$^-$, and XX on $U_{d300nm}$ is shown. For the calculation of $\tau$ the BDA method of Eq.~\ref{eq:CIOscStrengthPolWK} was used since it was shown in Fig.~\ref{fig:Lifetime}~(b) that it provides results more faithfully reproducing the experimental values for the studied weakly confined GaAs/AlGaAs QD system. We see in Fig.~\ref{fig:ELfldBindLife}~(c) that $\tau^{X}$ depends on $U_{d300nm}$ almost quadratically, increasing for both $U_{d300nm}<0$ and $U_{d300nm}>0$. Similar dependence on $U_{d300nm}$ around zero is seen also for X$^-$, albeit the values of $\tau^{X-}$ are $\sim 0.5$ smaller. Contrary to that, $\tau^{X+}$ and $\tau^{XX}$ show considerably asymmetric though mutually similar dependence on $U_{d300nm}$. For $U_{d300nm}<0$ the values of $\tau^{XX}$ and $\tau^{X+}$ first slightly reduce to $\tau\approx0.1$~ns and then increase for further decreasing $U_{d300nm}$ up to $\tau\approx0.2$~ns followed by a rapid increase of $\tau$, crossing the value of $\tau^{X}$ for $U_{d300nm}=-2.4$~V. On the other hand, for $U_{d300nm}>0$ $\tau^{XX}$ and $\tau^{X+}$ rapidly increase, reaching maximal values of $\tau^{XX}=2$~ns and $\tau^{X+}=1.25$~ns at $U_{d300nm}=1.9$~V and $U_{d300nm}=1.2$~V, respectively. A further increase of $U_{d300nm}$ leads to the reduction of $\tau^{XX}$ and $\tau^{X+}$ magnitudes towards the values of $\tau^{X}$. 
\begin{figure}[htbp]
    \includegraphics[width=85mm]{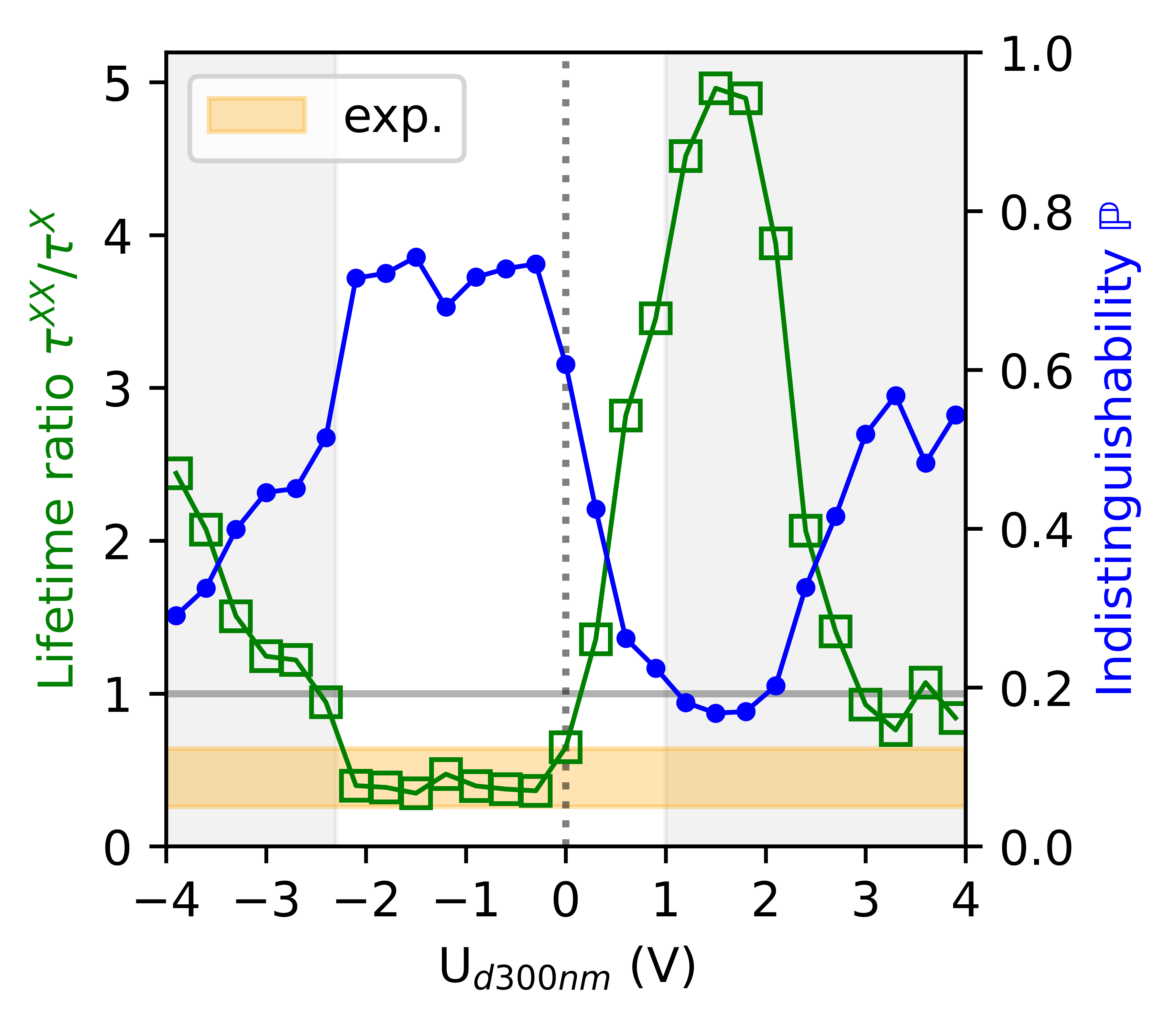}
	\caption{The ratio of XX and X$^0$ lifetimes, $\tau^{XX}/\tau^X$ from Fig.~\ref{fig:ELfldBindLife}~(c) is shown by green open squares. The photon indistinguishability $\mathbb{P}$ from Eq.~\eqref{eq:GabrielIndisting} is given by full blue balls. Orange shaded area marks the interval of $\tau^{XX}/\tau^X$ measured in Fig.~2~(d) of Ref.~\cite{Undeutsch2025}. The gray-shaded area correspond to voltages not considered in Ref.~\cite{Undeutsch2025}. The gray horizontal line marks $\tau^{XX}/\tau^X=1$,~i.e. the situation when lifetimes of X and XX are the same. In order to facilitate the comparison with Ref.~\cite{Undeutsch2025}, the electric field is given as a voltage applied on 300~nm thick layer, hence the label of horizontal axis of $U_{d300nm}$.}
	\label{fig:ELfldLifeRatio}
\end{figure}

The unusual behavior of XX and X$^+$ lifetimes can be explained by the different effective masses of electrons and holes, the former being much smaller than the latter as was discussed earlier. Since electrons are light, they do not feel the applied electric field that much as the holes which consist for all values of $U_{d300nm}$ of $>90$~\% of heavy holes. Hence, multi-particle complexes consisting of more than one hole, like XX and X$^+$ are influenced by $U_{d300nm}$ to larger extent. Conversely, in particular for X$^-$ the influence by $U_{d300nm}$ is rather timid.

The considerably smaller $\tau^{XX}$ than $\tau^{X}$ for $U_{d300nm}$ from $-2$~V to $0$~V was found advantageous in Ref.~\cite{Undeutsch2025} increasing the visibility of subsequently emitted photons by XX recombination in Hong-Ou-Mandel interference measurements. The indistinguishability of photons emitted in time domain is defined as~\cite{Undeutsch2025}
\begin{equation}
\label{eq:GabrielIndisting}
\mathbb{P} = \frac{1}{\frac{\tau^{XX}}{\tau^{X}}+1}.
\end{equation}
We show both $\frac{\tau^{XX}}{\tau^{X}}$ and $\mathbb{P}$ as a function of $U_{d300nm}$ in Fig.~\ref{fig:ELfldLifeRatio}. We compare our results of $\frac{\tau^{XX}}{\tau{X}}$, which we find for the interval of $U_{d300nm}$ from $-2$~V to $0$~V between $0.3$ and $0.45$, with measurements in Fig.2~d) of Ref.~\cite{Undeutsch2025} that are in the same voltage range between $0.3$ and $0.6$ (marked by orange shaded area in Fig.~\ref{fig:ELfldLifeRatio}). Thus, a surprisingly good agreement between theory and experiment is found. However, we note that for $U_{d300nm}$ in the range from $0$~V to $1$~V our results disagree with those in Ref.~\cite{Undeutsch2025} for the same interval. We attribute that disagreement to the fact that we used for our calculations a different QD than that which was measured in Ref.~\cite{Undeutsch2025} noting furthermore that in particular the emission properties of XX states are sensitive to QD properties and external perturbations~\cite{Bennett2005,Narvaez2005,Senellart2005,Alen2007,Undeutsch2025}.

Using Eq.~\eqref{eq:GabrielIndisting} we recalculate $\frac{\tau^{XX}}{\tau{X}}$ to indistinguishability $\mathbb{P}$ and show that by full blue balls in Fig.~\ref{fig:ELfldLifeRatio}. Clearly, the drop in $\tau^{XX}$ with respect to $\tau^{X}$ in the interval of $U_{d300nm}$ from $-2$~V to $0$~V is associated with $\mathbb{P}\approx0.75$ while for the rest of $U_{d300nm}$ we find $\mathbb{P}\approx0.2$ (except of the values of $U_{d300nm}$ from $3$~V to $4$~V when the electrons and holes are already considerably spatially separated by applied electric field and the emission of both types of complexes is fainter). Nevertheless, the calculations in this work confirm the large tunability of $\tau^{X}$ and $\tau^{XX}$ as well as their ratio.

\subsection{Role of preparation and detection of multi-particle states in GaAs/AlGaAs QDs}
\label{sec:Evgeny}
\begin{figure}[htbp]
    \includegraphics[width=85mm]{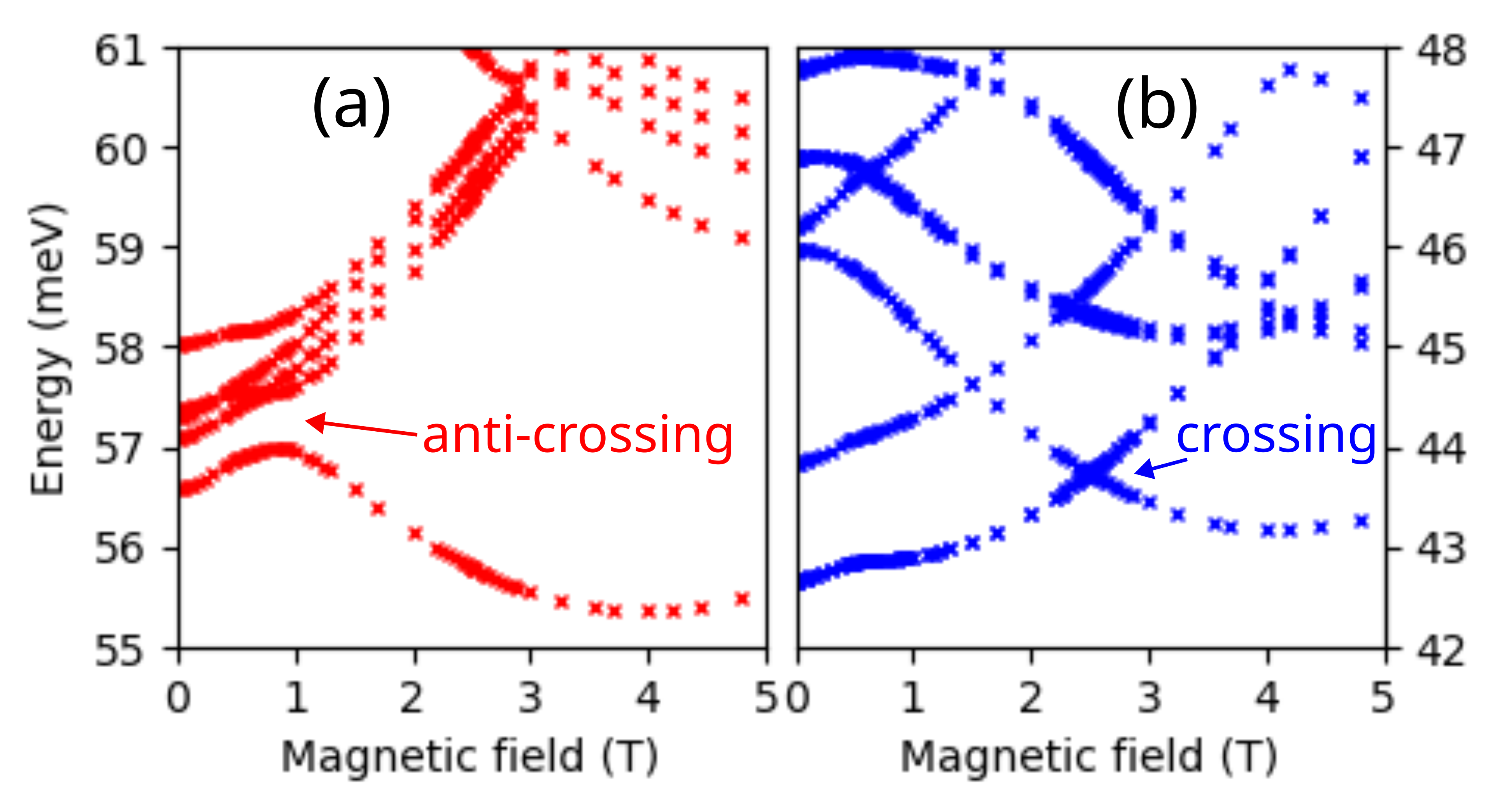}	
	\caption{Computed Coulomb interaction energies of the four electron states in GaAs QD as a function of the magnetic field applied along vertical QD dimension~\cite{Millington-Hotze2025}. The calculations in (a) [(b)] were done without [with] considering the electron-electron Coulomb exchange interaction. The data in (b) show a crossing of the singlet and triplet state for magnetic field around 2.5~T as previously measured in Ref.~\cite{Millington-Hotze2025}. On the contrary, data in (a) show only anti-crossing of singlet and triplet states. The four electron states in this figure were computed by CI with CI basis of ten single-particle electron states.}
	\label{fig:EC4eEn}
\end{figure}

To further study the role of the omission of the electron-electron and hole-hole exchange integrals, we now turn our attention to the $\mathbf{k}\!\cdot\!\mathbf{p}$~+~CI calculation of the complexes of interacting electrons which were experimentally studied in Ref.~\cite{Millington-Hotze2025}. There, with the help of the nuclear spin relaxation (NSR) measurements, it was found that the magnetic field applied on very similar GaAs/AlGaAs QDs as in this work caused a crossing of singlet and triplet states for the ground state of the complex of four interacting electrons. It is important to stress that the calculations in Ref.~\cite{Millington-Hotze2025} were performed exactly in the same fashion as here (including considering AFM QD structure exactly corresponding to the QDs in that paper,~i.e. slightly different than here) and with the same $\mathbf{k}\!\cdot\!\mathbf{p}$ and CI codes as in this work. We now repeat in Fig.~\ref{fig:EC4eEn} the calculations~\cite{Millington-Hotze2025} for the Coulomb energies of the four-electron complex in vertical magnetic field. In particular, we focus here on the results obtained without and with the inclusion of the Coulomb exchange between electrons, see Fig.~\ref{fig:EC4eEn}~(a)~and~(b), respectively. Clearly, for the calculation without electron-electron Coulomb exchange \{Fig.~\ref{fig:EC4eEn}~(a)\} no singlet-triplet crossing, observed in experiment~\cite{Millington-Hotze2025}, is found contrary to the calculation with Coulomb exchange \{Fig.~\ref{fig:EC4eEn}~(b)\}. Hence, the electron-electron Coulomb exchange interactions must not be omitted in those CI calculations to faithfully reproduce the NSR experiments. However, that is in contradiction to the results presented in Fig.~\ref{fig:BindingEXnoEEHH}~(b)~and~(d) where the omission of the electron-electron Coulomb exchange integrals (which have the largest magnitudes in Fig.~\ref{fig:BindingEXnoEEHH}~(c), even larger than hole-hole exchange) led to better agreement with PL experiments.

Since the multi-particle physics of the GaAs/AlGaAs QDs as well as their states must be qualitatively the same for both kinds of experiments, we conclude that it is the difference between how the multi-particle states are initialized and detected that necessitates a different theoretical treatment of calculating states in those experiments.

%
\section{Discussion}
\label{sec:discussion}
Finally, it is evident that the multi-particle calculations presented in this work, which involve omitting certain integrals to match the experimental results, lack elegance. However, even a fully self-consistent, correlated multi-particle solution would likely not fully capture the experimental observations in weakly confining QD systems. This is because, as demonstrated earlier, the theoretical description of results of multi-particle complexes observed in experiments depends on the specific conditions under which the system is prepared and measured. Concerning the former, whether the system is pumped using resonant~\cite{Undeutsch2025}, above-band excitation~\cite{Yuan2023}, or other methods (e.g. electric pumping~\cite{Millington-Hotze2025}). With respect to the latter, it is also important how the multi-particle states are probed, if it is by measuring their radiative emission~\cite{Yuan2023,Undeutsch2025} or interacting electrons and holes are studied via an interaction with some other system, like,~e.g., spins of atomic nuclei~\cite{Millington-Hotze2025}. We note that our XX calculations are compared to experiments in which XX was prepared by resonant two-photon excitation (TPE)~\cite{Schimpf2019,Undeutsch2025}, while the reference value from~\cite{DaSilva2021} originates from a perspective article that compiles results obtained under different excitation regimes. 
%

In summary, this underscores the fact that a comprehensive theoretical model describing the correlated multi-particle electronic structure of QDs would also need to properly account for the entire experimental setup, including the nature and effects of the excitation, followed by theory description of the time evolution of the multi-particle states including their possible interaction with environment (e.g.~phonons), and finally taking into account the properties of the detection setup.

{\color{black}We envisage that it may be insightful to test, experimentally and theoretically, how different probing schemes (e.g., above-band vs resonant PL, cathodoluminescence, and nuclear-spin-related techniques) affect the observed multi-particle complexes in the same GaAs/AlGaAs QD under otherwise identical conditions. Moreover, systematic size (volume) dependences of the energetic structure and radiative lifetimes, combined with correlated calculations, should provide further insight.}

%

\section{Conclusions}
\label{sec:conclusion}
We combined 8-band $\mathbf{k}\!\cdot\!\mathbf{p}$ model coupled to continuum elasticity with CI and a Poisson-based implementation of nonlocal (BDA) radiative rates to predict polarization-resolved oscillator strengths and lifetimes of X$^0$, X$^\pm$, and XX in weakly confining GaAs/AlGaAs quantum dots. The BDA calculation quantitatively matches independent lifetimes (e.g., $\tau^X\!\approx\!0.279\,\mathrm{ns}$, $\tau^{XX}\!\approx\!0.101\,\mathrm{ns}$) and reproduces electric-field trends, including the $\tau^{XX}/\tau^X$ controlled indistinguishability. We quantified sensitivity to CI basis and to exchange; in weak confinement, selectively omitting electron–electron and hole–hole exchange for specific complexes can improve agreement for PL observables, whereas other probes (e.g., nuclear spin relaxation spin spectroscopy) require exchange to recover level crossings. The workflow provides a reproducible route that connects realistic many-body wavefunctions with nonlocal light–matter coupling, and it can be extended to include preparation- and detection-specific kinetics (e.g., phonons, pure dephasing) relevant for device operation.



\section*{Acknowledgements}
\label{sec:acknowledgments}
%

The author thanks G.~Undeutsch, E.A.~Chekhovich, X.~Yuan, A.~Rastelli
for fruitful discussions and providing the experimental data.
The author acknowledges funding from the European Innovation Council Pathfinder program under grant agreement No 101185617 (QCEED),
support by the project Quantum materials for applications in sustainable technologies, CZ.02.01.01/00/22\_008/0004572, and partly funding by Institutional Subsidy for Long-Term Conceptual Development of a Research Organization granted to the Czech Metrology Institute by the Ministry of Industry and Trade of the Czech Republic.

\section*{Data availability}
\label{sec:DataAvailability}

The data that support the findings of this study are available from the corresponding author upon reasonable request.


%


\section{Appendix I.}
\label{sec:appendixI}
\begin{figure}[htbp]
	\includegraphics[width=85mm]{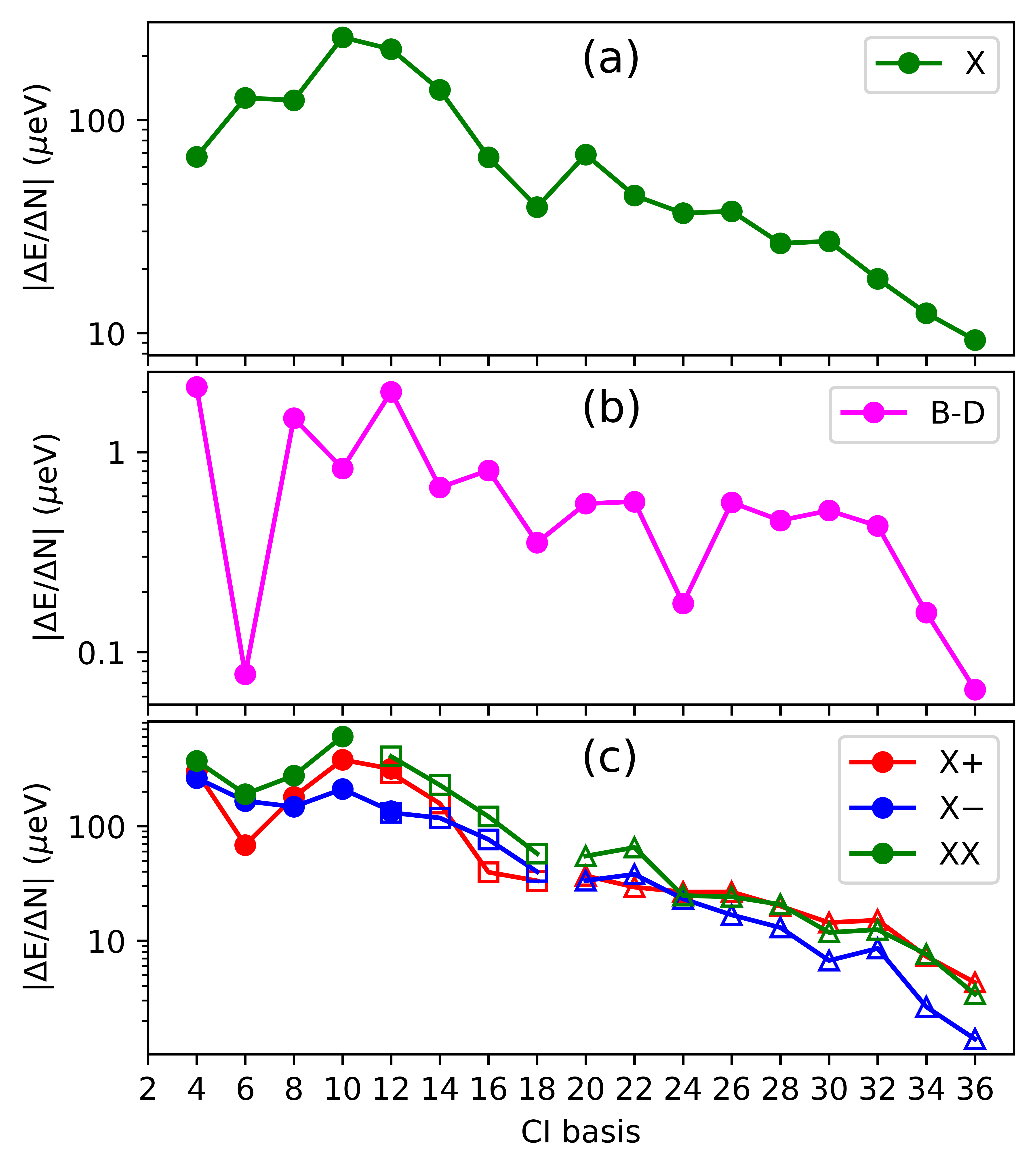}	
	\caption{We show the evolution of CI calculations for energies of (a)~X$^0$, (b)~bright-dark splitting, and (c)~X$^+$, X$^-$, XX binding with respect to X$^0$ as a function of the CI basis size. The dependencies are evaluated as an absolute value of the relative difference between energies ($E$) for consecutive CI basis state ($N$) as $|\Delta E/\Delta N|$. In each panel the left vertical axis is in logarithmic scale, hence an approximately linear decrease of $|\Delta E/\Delta N|$ for CI bases larger than $\sim 10$ in all panels is a clear sign of exponential nature of the convergence.}
	\label{fig:Econv}
\end{figure}

We show in Fig.~\ref{fig:Econv} the convergence study of the energies of X$^0$, bright-dark splitting, and the binding energies of X$^+$, X$^-$, XX with respect to exciton.


\section{Appendix II.}
\label{sec:appendixII}
\begin{figure}[htbp]
    \includegraphics[width=85mm]{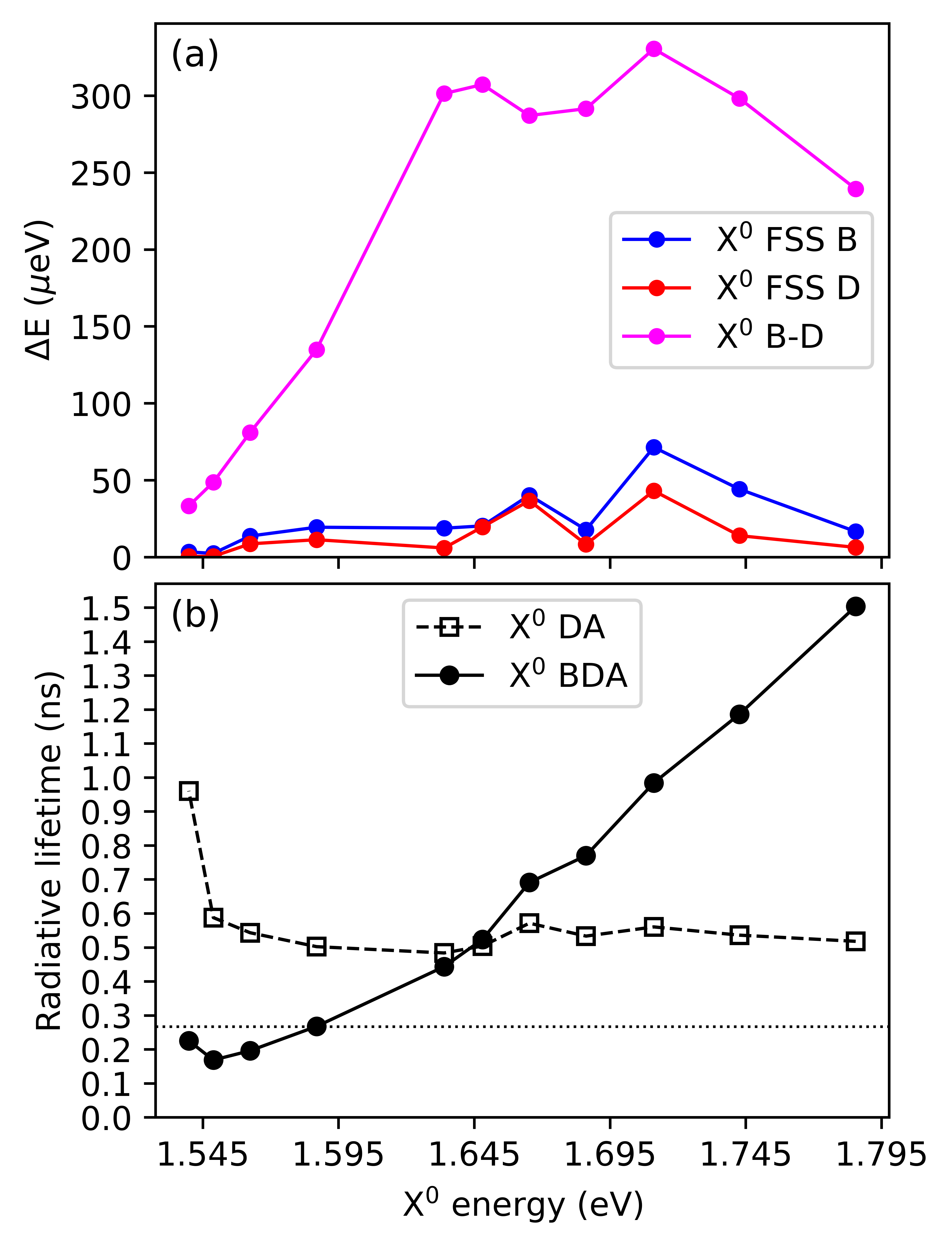}
	\caption{Calculations of volume dependencies of the multi-particle electronic and emission structure of cone shape GaAs QD in Al$_{0.4}$Ga$_{0.6}$As lattice, positioned on 2~nm GaAs layer,  similar (but not same) as that in Fig.~\ref{fig:AFMsp}~(a). We show in~(a)~bright (blue balls) and dark (red balls) X$^0$ FSS as well as bright-dark X$^0$ splitting (violet balls); in~(b)~the radiative lifetime of X$^0$ utilizing DA (empty squares) and BDA (full balls) method (see text)
    Note that the change of QD volume is identified on horizontal axes by X$^0$ energy. The largest X$^0$ energy (1.785~eV) corresponds to QD with basis diameter of $10$~nm and height of $2.5$~nm. On the other hand, the lowest X$^0$ energy (1.539~eV) correspond to dot with diameter of $70$~nm and height of $15$~nm.
    The horizontal black dotted line in (b) correspond to measured value of X$^0$ lifetime of 0.267~ns~\cite{Schimpf2019}.
    }
	\label{fig:LifeVdep}
\end{figure}
We show in Fig.~\ref{fig:LifeVdep} the evolution of the QD electronic and emission structure properties on QD volume. The calculations are performed for a cone-shaped GaAs QD in Al$_{0.4}$Ga$_{0.6}$As lattice \{different QD than that in Fig.~\ref{fig:AFMsp}~(a)\}, positioned on 2~nm GaAs layer (WL). The change of QD volume is achieved by fixing the QD aspect ratio (defined as height/diameter of QD) to 0.25 and varying the basis diameter from $10$~nm to $70$~nm. Using the aforementioned aspect ratio the latter change leads to the increase of QD height from $2.5$~nm to $15$~nm, respectively. In order to summarize the effect of QD volume change, we show the results in Fig.~\ref{fig:LifeVdep} as a function of the ground state exciton X$^0$ energy.

In Fig.~\ref{fig:LifeVdep}~(a) we give the QD volume evolution of bright and dark FSS as well as bright-dark energy splitting of X$^0$. We see that while both bright and dark FSS do not depend on QD size considerably, the bright-dark splitting seems more sensitive to GaAs QD volume. That might be the reason for the discrepancy of the computed B-D splitting in Fig.~\ref{fig:AFMsp}~(c) and measured value of $100\,\mu$eV~\cite{Yuan2023}.

In Fig.~\ref{fig:LifeVdep}~(b) we show the comparison of the evolution of emission radiative lifetime of X$^0$ for calculations that employed DA and BDA~\cite{Stobbe2012}. We clearly see the difference between DA and BDA approaches. Notably, apart of the largest dots (smallest X$^0$ energy), DA seems not to be much sensitive to QD volume. On the contrary, BDA leads to reduction of radiative lifetime with increase of QD volume up to QD with exciton energy of $1.5489$~eV upon which a further increase of QD volume leads to increase of radiative lifetime. The latter behavior is qualitatively similar to the calculations using DA method. Noticeably, for certain QD sizes (here for QDs emitting at $\sim 1.63$~eV), the DA and BDA approaches lead to similar emission lifetime of X$^0$. The aforementioned behavior was previously predicted in Ref.~\cite{Stobbe2012} being a general feature of the BDA method which is reproduced also in our calculations. The CI basis size for the aforementioned calculations was 36 single-particle electron and 36 single-particle hole states.

\end{document}